\documentclass[fleqn,10pt]{wlscirep}
\usepackage[utf8]{inputenc}
\usepackage[T1]{fontenc}
\usepackage{graphicx}
\usepackage{epsfig}
\usepackage{epstopdf}
\title{Second magnetization peak, anomalous field penetration, and Josephson vortices in KCa$_2$Fe$_4$As$_4$F$_2$ bilayer pnictide superconductor}

\author[1]{P. V. Lopes}
\author[1,*,$\dagger$]{Shyam Sundar}
\author[1,$\#$]{S. Salem-Sugui, Jr.}
\author[2,3,4]{Wenshan Hong}
\author[2,5]{Huiqian Luo}
\author[1]{L. Ghivelder}
\affil[1]{Instituto de Fisica, Universidade Federal do Rio de Janeiro, 21941-972 Rio de Janeiro, RJ, Brazil}
\affil[2]{Beijing National Laboratory for Condensed Matter Physics, Institute of Physics, Chinese Academy
of Sciences, Beijing 100190, China}
\affil[3]{School of Physical Sciences, University of Chinese Academy of Sciences, Beijing 100190, China}
\affil[4]{International Center for Quantum Materials, School of Physics, Peking University, Beijing 100871, China
}
\affil[5]{Songshan Lake Materials Laboratory, Dongguan, Guangdong 523808, China
}

\affil[*]{shyam.phy@gmail.com}
\affil[$\#$]{said@if.ufrj.br}

\affil[$\dagger$]{Presently at the School of Physics and Astronomy, University of St. Andrews, KY16 9SS, United Kingdom.}

\begin{abstract}
We performed magnetization measurements in a single crystal of the anisotropic bilayer pnictide superconductor KCa$_2$Fe$_4$As$_4$F$_2$, with $T_c$ $\simeq$ 34 K, for $H$$\parallel$$c$-axis and $H$$\parallel$$ab$-planes. A second magnetization peak (SMP) was observed in the isothermal $M(H)$ curves measured below 16 K for $H$$\parallel$$ab$-planes. A peak in the temperature variation of the critical current density, $J_{c}$($T$), at 16 K, strongly suggests the emergence of Josephson vortices at lower temperatures, which leads to the SMP in the sample. In addition, it is noticed that the appearance of Josephson vortices below 16 K renders easy magnetic flux penetration. A detailed vortex dynamics study suggests that the SMP can be explained in terms of  elastic pinning to plastic pinning crossover. Furthermore, contrary to the common understanding, the temperature variation of the first peak field, $H_1$, below and above 16 K, behaves non-monotonically. A highly disordered vortex phase, governed by plastic pinning, has been observed between 17 K and 23 K, within a field region around an extremely large first peak field. Pinning force scaling suggests that the point defects are the dominant source of pinning for $H$ $\parallel$$ab$-planes, whereas, for H $\parallel$$c$-axis, point defects in addition to surface defects are at play. Such disorder contributes to the pinning due to the variation in charge carrier mean free path, $\delta$$l$-pinning. Moreover, the large $J_c$ observed in our study is consistent with the literature, which advocates this material for high magnetic field applications.
\end{abstract}
\begin{document}

\flushbottom
\maketitle
%
\thispagestyle{empty}

\section*{Introduction}

Study of vortex dynamics and the investigation of different vortex-phases in superconductors are important for the fundamental understanding of vortex matter, as well as for technological advancement such as, in next generation particle accelerator technology. The discovery of the iron-pnictide superconductors renewed the interest in exploring the vortex phase diagram in superconductors \cite{1}. This is due to the fact that these materials possess intermediate to high superconducting transition temperature, $T_c$ \cite{Tc,stewart}, high upper critical magnetic fields, $H$$_{c2}$ \cite{4, 5, jaro, senatore, 3}, high  critical current density, $J_c$, sustained up to considerable large applied magnetic fields \cite{4,5}, a better intergrain connectivity when compared to the cuprates and nickelates\cite{katase,durrell,13,14}, and low anisotropy \cite{3,gama1}. Also, the pnictides have multiband pairing with possible implications for vortex pinning, as carriers are interband or intraband scattered \cite{multi}. As a result of these properties, many vortex pinning studies, including vortex dynamics, have been developed in pnictides looking towards  technological applications \cite{ren,app1,app2,app3,app3a,app4,app5,app6,hosono}. Among the features present in the irreversible region of many iron-pnictide superconductors, the second magnetization peak, SMP, appearing in isothermal magnetization curves, is the most studied. In addition to the basic understanding of the SMP, this feature is also interesting due to the fact that it renders a peak in the critical current density in the same magnetic field range where SMP appears in $M(H)$. This makes the SMP directly related to a technologically relevant parameter, the critical current density. The SMP has been observed in most of the iron-pnictide superconductors, with magnetic fields applied parallel and perpendicular to the $c$-axis of the crystals, but mostly studied with $H$$\parallel$$c$-axis, and its origin has been shown to be system dependent, with no consensus so far for its actual cause. The proposed explanations for the SMP in iron-pnictides are, a softness of the vortex lattice \cite{baruch,baruch1,soft}, a pinning crossover from elastic to plastic \cite{app5,15,16,17,18,19,20,21}, an order-disorder transition \cite{order-disorder1,order-disorder2,order-disorder3}, and a vortex-lattice phase transition \cite{baruchII,baruch1,baruch2,VLPT1,VLTP2,miu,VLTP3}. Moreover, a peak in the magnetic field dependence of the relaxation rate has been attributed to a precursor mechanism that leads to a SMP at higher fields due to a crossover from a low effective pinning to a strong effective pinning \cite{Polichetti:2021}. On the other hand, the explanation for the disappearance of the SMP is based on a plastic pinning replacing collective pinning as temperature increases \cite{ep5}. The SMP has been previously observed and studied in the high-$T_c$ cuprates mostly for $H$$\parallel$ $c$-axis \cite{abulafia,kwok,bisco,eley,cole}, as well as in the conventional low-$T_c$ superconductors \cite{low}. It is worth mentioning that most of the vortex dynamics studies on the SMP found in the literature were conducted in less anisotropic iron-pnictides systems, with few studies on systems with moderated anisotropy (112-family and 1111-family) (with $\gamma$$\sim$8)\cite{mod1,mod2,17}, which also shows a SMP for $H$$\parallel$$c$-axis. 

In this work, we explore the vortex dynamics in a newly discovered anisotropic bilayer pnictide superconductor, KCa$_2$Fe$_4$As$_4$F$_2$ (ref. \cite{Ni=0}) where the FeAs layers are alternately separated by conductive K layers and insulating CaF$_2$ layers, which yield a large anisotropy, $\gamma$$\sim$15, close to $T_c$ (ref. \cite{wang, pin1}). Previous vortex pinning studies on KCa$_2$Fe$_4$As$_4$F$_2$ (ref. \cite{pin1,58,pin2}) did not study vortex dynamics or the existence of the SMP in this system, which is addressed here. We study a high quality single crystal with $T_c$ $\simeq$ 34 K, in which vortex dynamics revealed a distinct and exotic behavior when compared to other iron-pnictide systems. Our study was performed with the magnetic field applied both parallel and perpendicular to the $ab$-planes of the sample. Isothermal magnetic field dependence of magnetization, $M(H)$, obtained for $H$$\parallel$$ab$-planes, show a SMP which develops only at temperatures below 16 K. Contrary to reports on other iron-pnictide superconductors, the SMP in our sample is absent for $H$$\parallel$$c$-axis, and only develops for $H$$\parallel$$ab$-planes. The observed SMP appears to be directly associated to a peak in the isofield temperature variation of the critical current density, $J_c$($H$ = 0, $T$) and $J_c$($H$ = 10 kOe, $T$), at 16 K for $H$$\parallel$$ab$-planes. We argue that such a peak in $J_c$($H$ = 0, $T$) and $J_c$($H$ = 10 kOe, $T$) is associated to the large anisotropy of the sample, with the consequent emergence of Josephson vortices occurring below 16 K, and producing the SMP. An analysis based on activation energy, $U(M)$, suggests that this SMP can be explained in terms of a crossover from elastic pinning to plastic pinning. We also observed the development of an anomalous first peak in the isothermal $M(H)$ curves for $H$$\parallel$$ab$-planes at temperatures above 16 K and below 23 K. We noticed that the magnetic field associated to the first peak, $H_1$, in the temperature range 16-23 K, is much larger than the $H_1$ found below 16 K and above 24 K. Also, the first peak in $M(H)$ curves for $H$$\parallel$$ab$-planes at temperatures above 23 K, are of the same values as those found below 16 K for which the SMP is present. This fact evidences that the magnetic field penetrates easily below 16 K, which appears to be related to the emergence of Josephson vortices. Our vortex dynamics study performed for $H$$\parallel$$ab$-planes tries to address the origin of this intriguing extremely large first peak field (called $H_a$) appearing in the $M(H)$ curves for $H$$\parallel$ $ab$-planes above 16 K and below 23 K. This experimental evidence currently lacks an explanation, and to the best of our knowledge it has not been observed before in any system. On the other hand, the study performed for $H$$\parallel$$c$-axis shows $J_c$(0) exceeding 10$^6$ A/cm$^2$ at $T$ $\leq$ 14 K, which signals that the system has potential for technological applications. The scaled pinning force curve obtained for $H$$\parallel$$c$-axis suggests the effective role of point pinning along with surface pinning. On the other hand, the scaling of the pinning force curves for $H$$\parallel$$ab$-planes suggests the point defects as the dominant source of pinning. 

\section*{Results \& Discussion}

\begin{figure}[htb]
\includegraphics[width=\linewidth]{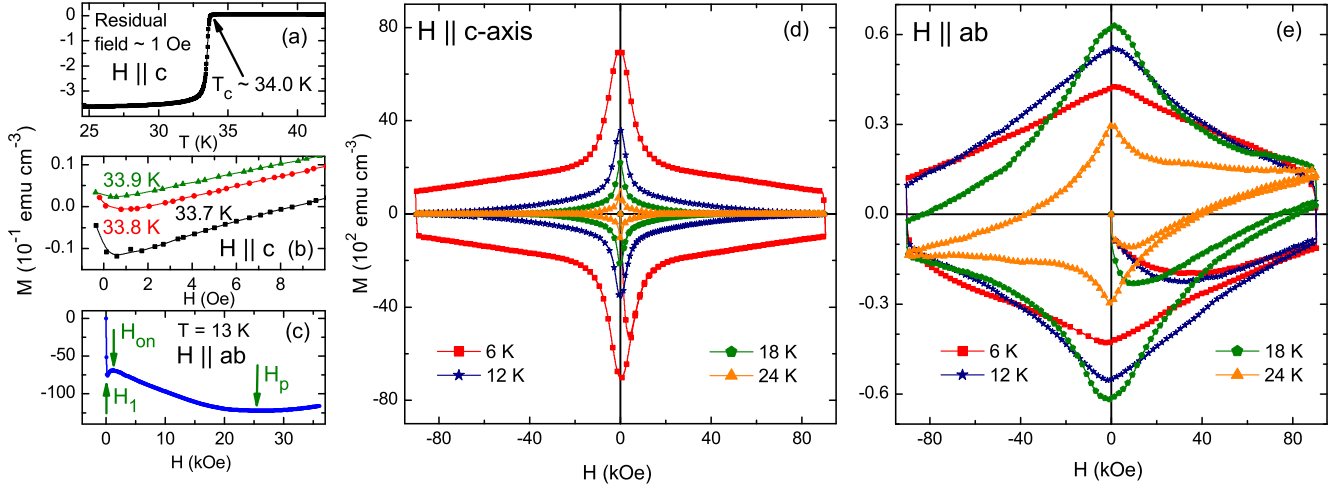}
\caption{(a) Temperature dependence of magnetization measured with $\sim$1 Oe magnetic field. The arrow indicates the onset of superconducting transition, $T_c$$\sim$34 K. (b) Isothermal magnetic field dependence of magnetization, $M(H)$, measured at temperatures close to $T_c$. (c) Initial branch of the isothermal $M$($H$) measured at 13 K shows the signature of SMP. Arrows indicate the characteristic penetration field ($H_1$), onset field ($H_{on}$) and peak field ($H_p$), associated to the first and second magnetization peaks in isothermal $M(H)$. (d, e) Representative isothermal $M(H)$ curves measured at various temperatures below $T_c$ for $H$$\parallel$$c$-axis, and for $H$$\parallel$$ab$-planes.}
\label{fig1}
\end{figure}

Figure 1 shows selected isothermal magnetization curves, $M(H)$, obtained for $H$$\parallel$$c$-axis and $H$$\parallel$$ab$-planes. Each $M(H)$ data was obtained by cycling the magnetic field from positive to negative and again to positive values corresponding to 5 field branches, as shown in the curves of Fig.1. In fig. 1a, the temperature dependence of magnetization, $M(T)$, measured in the zero field cooled (ZFC) state with $H$ $\sim$ 1 Oe, shows $T_c$ $\sim$ 34 K, and $\Delta$$T_c$ = 0.7 K. Figure 1b shows isothermal $M(H)$ curves obtained at temperatures just below $T_c$, evidencing the high quality of the sample. Figure 1c exhibits the details of the SMP at 13 K showing the onset field, $H_{on}$, and the peak field, $H_p$. Figure 1d shows that all $M(H)$ curves for $H$$\parallel$$c$-axis are symmetric relative to the $x$-axis, indicating that bulk pinning is dominant in the sample. However, for $H$$\parallel$$ab$-planes, asymmetric $M(H)$ curves were observed for temperatures close to $T_c$. This might be due to the large anisotropy near $T_c$. The SMP is only observed for $H$$\parallel$$ab$-planes at temperatures below 16 K, and only in the first branch of the isothermal $M(H)$. The upper branch of the $M(H)$ curves for $H$$\parallel$$ab$-planes below 16 K do not follow the trend observed in the $M(H)$ curves above 16 K. The increase in magnetization as the field decreases is comparatively lower in the curves below 16 K when compared with the curves obtained above 16 K. As shown below, this produces a peak in the isofield $J_c($T$)$ at 16 K. From the $M(H)$ curves in Fig.1e, for $T$ $\leq$ 16 K we can extract the first peak field, $H_1$, associated with field penetration in the sample, the onset field $H_{on}$ above which the SMP develops, the SMP peak field $H_p$ and the irreversible field $H_{irr}$. The values of $H_{irr}$ are obtained as the point where the 1$^st$ and 2$^nd$ branches of curves in each $M(H)$ merge together (see figure S1 in supplementary materials). 

Figure 2a shows the increasing field branch of $M(H)$ curves measured at temperatures in the range 21-26 K for $H$$\parallel$$ab$-planes. The figure helps to show the behavior of the first peak field $H_1$ for $H$$\parallel$$ab$-planes, associated to field penetration. Figure 2a clearly shows the usual increase of $H_1$ down to 24 K, however, an abrupt enhancement in $H_1$ is observed as the temperature drops below 24 K. Moreover, for temperatures in the range 23-17 K, $H_1$ values are similar to the SMP peak field $H_p$. For this reason we call the first peak field at temperatures between 23 K and 17 K as $H_a$, since the first peak field in this temperature region is too large to be interpreted as the penetration peak field. An abrupt increase of penetration field from 480 Oe to 2170 Oe in the small temperature window of 24 K to 23 K respectively, is clearly demonstrated in the inset of Fig. 2a. It is important to notice that as temperature further decreases below 16 K, where the SMP appears, the first peak $H_1$ has a much lower value than $H_p$ as evidenced in Fig.1c, with values of the same order of magnitude of $H_1$ in the temperature region above 24 K (also see fig. 2b).

\begin{figure}[htb]
\includegraphics[width=\linewidth]{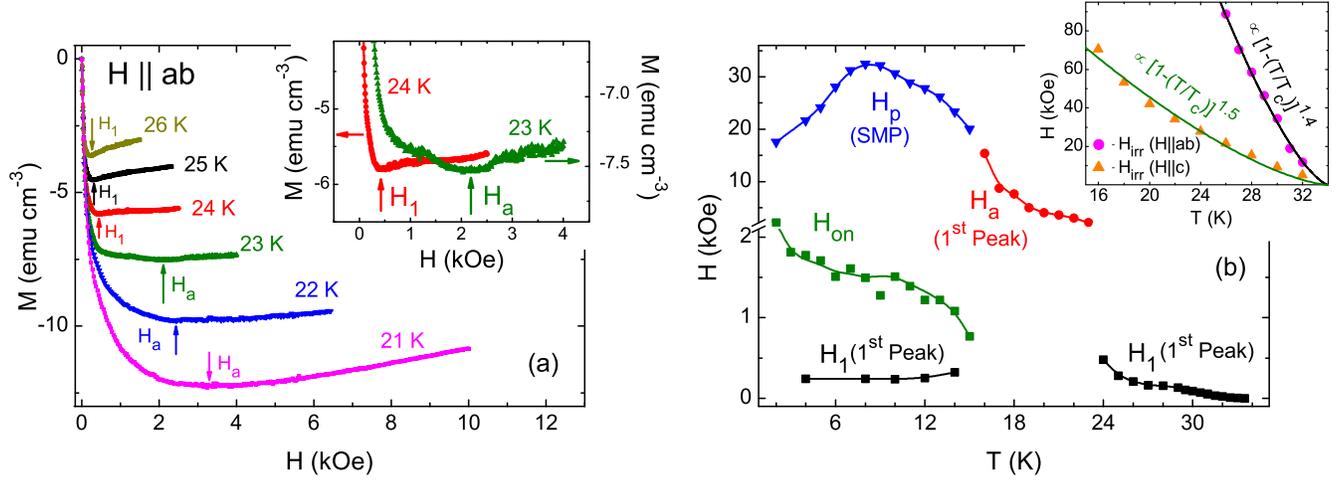}
\caption{(a) Initial branch of the isothermal $M(H)$ curves measured between 21 K and 26 K. The arrows indicate the observed first peak in each $M(H)$ curve. The inset evidences an abrupt enhancement in the first peak field value at 23 K compared to the one observed at 24 K. (b) Temperature dependence of magnetic field associated to the marked changes observed in each isothermal $M(H)$ measured for $H$$\parallel$$ab$-planes (see text for details). The lines are guide to the eyes.  Inset shows the irreversibility lines obtained for $H$$\parallel$$c$-axis, and for $H$$\parallel$$ab$-planes.}
\label{fig2}
\end{figure}

Figure 2b shows the phase diagram obtained from the $M(H)$ curves for $H$$\parallel$$ab$-planes with the values of $H_1$, $H_a$, and $H_p$. In the inset of Fig.2b, we show $H_{irr}$ obtained for $H$$\parallel$$ab$-planes and $H$$\parallel$$c$-axis. We observed that the temperature dependence of $H_{irr}$($T$) follows the expression (1-$T$/$T_c$)$^{1.4}$ and (1-$T$/$T_c$)$^{1.5}$ for $H$$\parallel$$ab$-planes and $H$$\parallel$$c$-axis respectively. Similar temperature dependence has also been seen in case of YBa$_2$Cu$_3$O$_{7-x}$ (ref. \cite{il}). The zero field limit of the irreversible field, H$_{irr}$(0), for $H$$\parallel$$ab$-planes and $H$$\parallel$$c$-axis are 70 kOe, and 16.7 kOe. Consequently, the zero field limit of the anisotropy parameter of irreversible field, $\gamma_{Hirr}$ = $H_{irr, ab}$/$H_{irr, c}$, is found to be $\sim$ 4. Interestingly, this is in good agreement with ref. \cite{wang}, where it is shown that the anisotropy parameter of the upper critical field, $\gamma_{H_{c2}}$, decreases as temperature goes below $T_c$, and achieves $\gamma$ $<$ 5 below 26 K. Here, it is to note that, since the anisotropy parameter obtained from $H_{c2}$ is closely related to the anisotropy of the effective electron mass, whereas the irreversible field can be influenced by the flux-pinning properties including flux-creep effects, therefore, a comparison of anisotropy parameters obtained from $H_{c2}$ and $H_{irr}$ may not be appropriate for temperatures near $T_c$, where the flux-creep effects are prominent. However, in case of iron-pnictide superconductors, it has been observed that the anisotropy parameter of upper critical field ($\gamma_{H_{c2}}$) monotonically decreases with decreasing temperature \cite{Zhang:2011}, and for KCa$_2$Fe$_4$As$_4$F$_2$ (ref. \cite{wang}) it saturates with $\gamma_{H_{c2}}$ $<$ 5 at low temperatures. Therefore, in the zero temperature limit, it seems reasonable to compare the anisotropy parameter obtained from the ${H_{c2}}$ and $H_{irr}$, if the dominant pinning is isotropic in nature, such as pinning due to the point defects. Comparable values of the anisotropy parameters obtained from the upper critical field and irreversible field, $\gamma_{H_{c2}, H_{irr}}$ $\sim$ 3, have been realized in a BaFe$_{1.9}$Ni$_{0.1}$As$_2$ iron-pnictide superconductor \cite{soft, Tao:2009, Rey:2013}. Figure 2b clearly shows the discontinuous change in the first peak as $T$ drops below 24 K, also visible in Fig.2a and its inset. The values of $H_a$ at temperatures just above 16 K are of the same magnitude as $H_p$, occurring below 16 K. In fact, the $H_a$ line smoothly joins the $H_p$ line. Therefore, one may speculate that $H_a$ might be associated to a phenomenon related to the SMP, which develops without the appearance of the onset field, $H_{on}$. It is also possible to see that the values of $H_1$ at temperatures above 24 K are of the same magnitude as $H_1$ below 16 K, for which the SMP develops. By looking at the behavior of $H_1$ as $T$ drops below $T_c$ one would expect large values of $H_1$ below 16 K. As discussed below, the considerably low values of $H_1$ below 16 K appear to be associated to the formation of Josephson vortices, evidencing that the magnetic field penetrates easily when Josephson vortices form inside the sample, instead of Abrikosov vortices. The $H_p$ line in Fig. 2b shows a peak at 8 K, with $H_p$ decreasing as temperatures decreases below 8 K. This behavior is quite different from the usual increasing of $H_p$ as temperatures decreases, observed in other systems, and seems to be related to the Josephson vortices in the system. To our knowledge, the KCa$_2$Fe$_4$As$_4$F$_2$ system is the first pnictide superconductor, which, presents the SMP only for $H$$\parallel$$ab$-planes, exhibits the anomalous peak field $H_a$, and also shows non-monotonic temperature dependence of the field associated to magnetic flux penetration. A similar SMP appearing only for $\parallel$$ab$-planes was recently observed for the electron doped cuprate superconductor, Pr$_{0.87}$LaCe$_{0.13}$CuO$_4$, with an anisotropy $\gamma$$\sim$ 10, and the SMP was explained in terms of an elastic pinning to plastic pinning crossover \cite{us}. The SMP in that case was observed for $T$ $\le$ 0.65 $T_c$, including the low temperature region where the Josephson vortices were observed. A SMP appearing only for $\parallel$$ab$-planes was also observed in La$_{2-x}$Sr$_x$CuO$_4$ in the doping domain of static charge and spin stripes, however, for $H$$\parallel$$c$-axis no SMP was noticed\cite{miu2}.

To address the mechanism associated with the appearance of the SMP for $H$$\parallel$$ab$-planes as well the appearance of the anomalous first peak, $H_a$, above 16 K, we performed magnetic relaxation measurements, $M(t)$, for field values below and above the SMP field $H_p$ and $H_a$, on selected isothermal $M(H)$ curves with $H$$\parallel$$ab$-planes. A representative $M(H)$ curve, with magnetic relaxation measured on its initial branch, is shown in figure S2 in the supplementary materials. We also measured $M(t)$ curves as a function of temperature for fixed magnetic fields for $H$$\parallel$$ab$-planes. The $M(t)$ curves were also obtained as a function of $H$ over selected isothermal $M(H)$ curves and as a function of temperature at fixed magnetic fields for $H$$\parallel$$c$-axis. All logarithmic ln$M(t)$ vs. ln$t$ curves for $H$$\parallel$$ab$-planes were found non-linear, while, the usual linear behavior in ln$M$ vs. ln$t$ curves for $H$$\parallel$$c$-axis allowed us to obtain the relaxation rate $R$ = dln$M$/dln$t$ (see figures S3 and S4 in the supplementary materials). Each $M(t)$ curve for both directions yield a respective activation energy curve $U(M)$ \cite{maley}. The resulting figures, $R$ vs. $H$ and $R$ vs. $T$ curves for $H$$\parallel$$c$-axis displays a continuous increase with field and temperature, which is usually observed due to enhanced flux creep caused by increased vortex density with field and increased thermally activated flux creep with temperature (see figure S4 in the supplementary materials).  

\begin{figure}[htb]
\includegraphics[width=\linewidth]{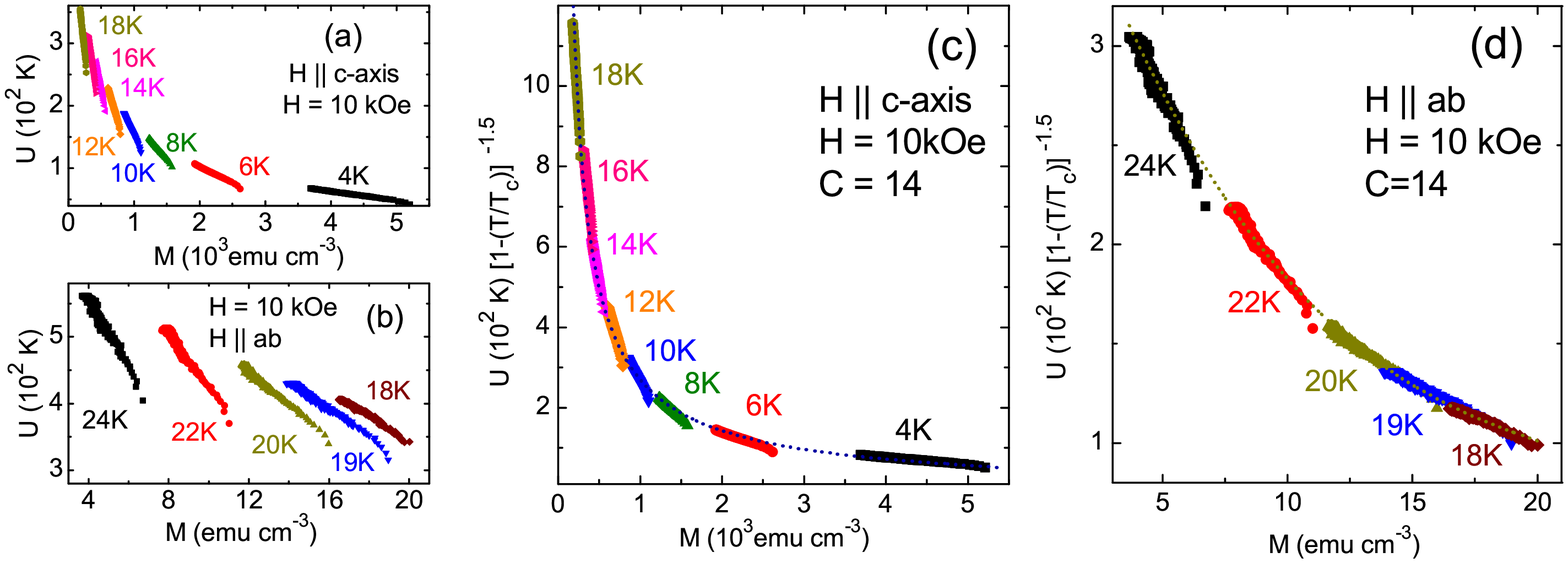}
\caption{(a, b) Magnetic field dependence of activation energy, $U(M)$ = -$T$lnd$M$/d$t$ + $14$$T$, obtained from magnetic relaxation data measured with $H$ = 10 kOe at various fixed temperatures for $H$$\parallel$$c$-axis, and for $H$$\parallel$$ab$-planes. (c, d) Scaling of $U(M)$ curves using (1-$T$/$T_c$)$^{1.5}$ function is achieved for both field orientations.}
\label{fig3}
\end{figure}

Figures 3a and 3b show the activation energy $U(M)$ obtained with a fixed field $H$ = 10 kOe as a function of temperature for both field directions, with $U(M)$ = -$T$lnd$M$/d$t$ + $C$$T$, where $C$ = 14 is a constant which depends on the attempt frequency, hoping distance and sample dimensions \cite{maley}. The $U(M)$ curves for a fixed $H$ at different fixed temperatures  is expected to be a functional form  of $M$ after scaling with a temperature function, $(1-T/T_c)^{1.5}$ (ref. \cite{maley, maley2}). This procedure allowed us to determine the value of $C$ = 14 for $H$$\parallel$$c$-axis and for $H$$\parallel$$ab$-planes at temperatures above 16 K, as shown in Fig.3c and 3d. However, we could not find such a functional form of $U(M)$ for $H$$\parallel$$ab$-planes for $T$ below 16 K (see figures S5 an S6 in the supplementary materials). The latter is possibly related to the emergence of Josephson vortices below 16 K, which do not have a normal core \cite{JV1}. As discussed below, for $H$$\parallel$$ab$-planes below 16 K, a Josephson vortex lattice takes place. Since it is expected that the same value of $C$ holds in the entire range of temperatures and fields, we obtained the corresponding $U(M)$ curves from $M(t)$ measured along the isothermal $M(H)$  exhibiting the SMP for $H$$\parallel$$ab$-planes. 

Figure 4a shows the resulting $U(M)$ curves obtained for the isothermal $M(H)$ curve at $T$ = 4 K, and at field values below and above the peak field $H_p$. The clear difference in the behavior observed on the $U(M)$ curves as $H$ crosses the SMP is characteristic of a pinning crossover. Figure 4b shows that each different set of $U(M)$ appearing in the top panels can be scaled as $U(M)$$H^{-n}$ with $n$ = 1.2 for $H$ below $H_p$ and $n$ = -1 for $H$ above $H_p$, in agreement with the behavior expected for an elastic pinning (collective) to plastic pinning crossover \cite{abulafia, Burlachkov:2022}. Figure 4c shows $U(M)$ curves calculated with $C$ = 14 obtained from the $M(t)$ curves for selected field values lying below and above the peak field $H_a$, on the $M(H)$ at 20 K for $H$$\parallel$$ab$-planes. Figure 4c shows that the $U(M)$ curves obtained for magnetic fields below and above the anomalous peak field $H_a$ do not show any considerable change in the behavior as the peak field $H_a$ is crossed. We observed in Fig. 4d that the $U(M)$ curves shown in Fig. 4c follows the same $U(M)$$H$$^{-n}$ scaling behavior, but with an exponent $n$ = -0.5 above $H_a$ and $n$ = -0.1 below $H_a$. The exponent $n$ = -0.5 obtained above $H_a$ is close to the one expected for plastic pinning, while the exponent $n$ = -0.1 obtained below $H_a$ is too small, but shows the same trend with $H$ associated to plastic pinning \cite{abulafia, Burlachkov:2022}. The increase of magnetization with increasing field, observed below $H_a$, is not expected to be compatible with plasticity, therefore it is likely that some other vortex pinning mechanism associated with plasticity produces the increase of magnetization with field. In the literature, plastic pinning for fields below the $H_p$ has been observed in case of Nd$_{1.85}$Ce$_{0.15}$CuO$_{4-\delta}$ and YBa$_2$Cu$_3$O$_{7-\delta}$ superconductors \cite{Giller1997, Giller1999}. Such plastic pinning for $H$$<$$H_p$ can be realized in the disordered vortex lattice which appears at low fields near $H_{on}$. It is important to note that below $H_p$ the plastic pinning only dominates at high enough fields, where vortex lattice has significant disorder. Therefore, due to the competition between energy associated to the plastic pinning and elastic pinning, the observed scaling exponents might be different than the one theoretically expected. However, $n$ = -0.1 below $H_a$ at 20 K strongly suggests the plastic nature of vortex pinning in the present case. Although the exact reason for the absence of $H_1$ and $H_{on}$ in the $M(H)$ curves above 16 K and below 23 K is unknown, but it might be associated to the plasticity of the vortex lattice near the full field penetration. Since the SMP only appears for $H$$\parallel$$ab$-planes, it is likely that intrinsic pinning associated to the layered structure of the system plays a significant role in the existence of the SMP \cite{us}.

\begin{figure}[htb]
\includegraphics[width=\linewidth]{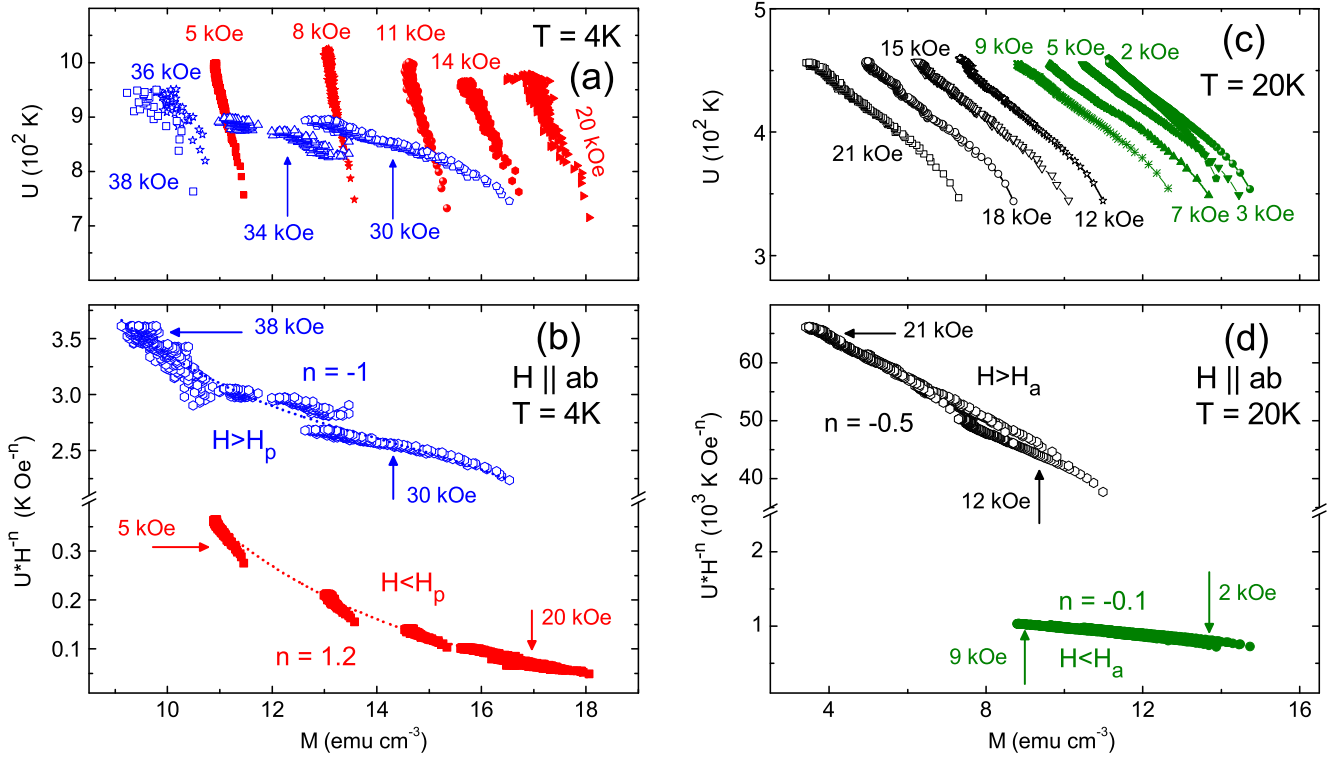}
\caption{(a, c) Activation energy as a function of magnetic field, $U(M)$, measured over isothermal $M$($H$) across the peak field $H_p$ and $H_a$ at 4 K and 20 K respectively. (b) Scaling of $U(M)$ curves with $H^n$ suggests elastic to plastic pinning crossover across $H_p$. (d) Scaling of $U(M)$ curves with $H^n$ indicates plastic pinning over the whole field range for $T$ = 20 K. It suggests a disordered vortex lattice as the field penetrates the sample. Smaller value of the exponent $n$ at lower fields might be due to the competing nature of energy for elastic and plastic pinning at low fields (see details in the text).}
\label{fig4}
\end{figure}

Figure 5 shows the magnetic field dependence of the critical current density, $J_c(H)$, obtained from the isothermal $M(H)$ data using the Bean's critical state model \cite{bean}, with $J_c$ = 20 $\Delta$$M$/$a_1$(1-$a_1$/3$b_1$), where, $b_1$ $>$ $a_1$ (units in cm) are the dimensions of the single crystal defining the area perpendicular to the applied magnetic field, $\Delta$$M$ (emu/cm$^3$) is the width of the $M(H)$ curves obtained by subtracting the 2$^{nd}$ from 5$^{th}$ branch of the $M(H)$ curves, with the resulting $J_c$ given in A/cm$^2$. Note that for $H$$\parallel$$ab$-planes, $a_1$ corresponds to the thickness of the sample. The value of the critical current density for zero magnetic field, $J_c$($H$ = 0), for $H$$\parallel$$c$-axis, is rather large, exceeding 10$^6$ A/cm$^2$ at temperatures below 14 K. Large $J_c$ observed in our work is consistent with the recent literature \cite{Pyon:2020}, evidencing that the system has potential for technological applications. Since $J_c$ is calculated by subtracting the 2$^{nd}$ from 5$^{th}$ branch of the $M(H)$ curves, which do not show the SMP, the peak effect associated to the SMP does not appear in the $J_c(H)$ curves below 16 K. Following the usual trend, $J_c$($H)$ for $H$$\parallel$$c$-axis continuously decreases as the field increases forming a downward curvature curve. For $H$$\parallel$$ab$-planes, $J_c$($H$) curves shows an inflection point at low fields which separates a downward curvature curve for low fields from an upward curvature as the field increases. Another feature visible for $H$$\parallel$$ab$-planes is that the $J_c$($H$) values at temperatures below 16 K, for fields below 20 kOe, are lower than the same $J_c$($H$) values for the curves at 16 K and 18 K. For instance, the value of $J_c$($H$) below 20 kOe at 3 K, 6 K, and 10 K are lower than the $J_c$($H$) below 20 kOe at 16 K and 18 K. This feature can be better seen by plotting the temperature dependence of the isofield critical current density, $J_c($H$ = 0, $T$)$, and $J_c$($H$=10 kOe, $T$), for $H$$\parallel$$ab$-planes, as shown in Fig. 5c. A clear peak in $J_c$($T$) at 16 K is observed. It is highly suggestive that this peak in $J_c(T)$ is directly related to the appearance of the SMP occurring below 16 K for $H$$\parallel$$ab$-planes. Two possible explanations for such a drop in $J_c$($T$) below 16 K are: an abrupt change in the volume pinning which is discarded, or a vortex-lattice phase transition, which includes a change in the vortex-matter associated to a dimensional 3$D$-2$D$ crossover, with 3$D$ Abrikosov vortices giving place to 2$D$ Josephson vortices below 16 K. The scenario of a dimensional crossover in very anisotropic layered systems is possible, since the coherence length decreases with temperature until the emergence of Josephson vortices lattice is favorable \cite{JV0}. The transition of an Abrikosov vortex lattice to a Josephson vortex lattice should be followed by a change in the magnetic flux inside the sample \cite{JV0} which was observed in the $M(H)$ curves of Fig.1e below 16 K. A further explanation for the drop in $J_c$($T$) being associated to such a change in the vortex matter relies on the differences between Abrikosov and Josephson vortices. While Abrikosov vortices have two characteristics length scales, the normal core of the size of the coherence length and the London penetration depth $\lambda$, Josephson vortices do not have a core and have only one length scale given by $\lambda$$_j$ = ($c$$\phi$$_0$16$\pi$$^2$$\lambda$$j_s$)$^2$, where $\phi$$_0$ is the quantum flux, $c$ is the velocity of the light, $\lambda$ is the London penetration depth and $j_s$ is the Josephson critical current density, which is smaller than $J_c$ \cite{JV1}. Since $\lambda$$_j$ is usually much higher than $\lambda$, Josephson vortices are more weakly pinned than Abrikosov vortices \cite{JV1,JV2} supporting that the emergence of Josephson vortices occurring below 16 K would produced a drop in $J_c$($T$), as observed in Fig. 5c. A similar peak in $J_c$($T$), but not followed by a SMP, was observed in SmFeAs(O,F) with $T_c$$\sim$ 48-50 K and $\gamma$$\sim$ 4-6. The authors claimed the peak is a consequence of well pinned slow moving Abrikosov vortices at high temperatures changing to weakly pinned fast moving Josephson vortices at low temperatures \cite{eur}. A similar peak in $J_c$($H$ = 0, $T$) for $H$$\parallel$$ab$-planes was observed more recently in Pr$_{0.87}$LaCe$_{0.13}$CuO$_4$ and explained in terms of a dimensional 3$D$ to 2$D$ crossover with Abrikosov vortices giving place to Josephson vortices as temperature is lowered. In the latter, the SMP was visible in the  temperature region below and above the peak in $J_c$(T) for $H$$\parallel$$ab$-planes. In the present study, it is clear that the SMP only appears below 16 K associated to the emergence of Josephson vortices. 

\begin{figure}[htb]
\includegraphics[width=\linewidth]{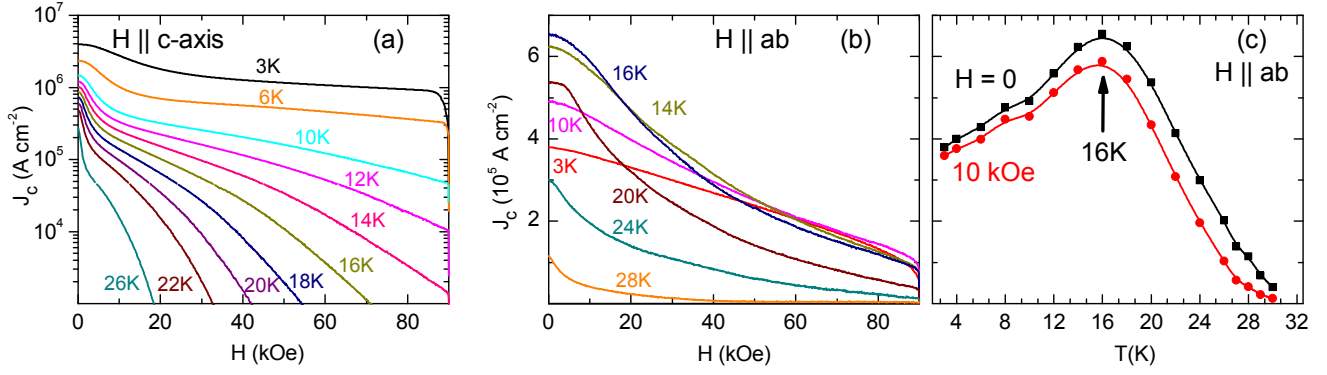}
\caption{(a,b) Magnetic field dependence of critical current density, $J_c(H)$, at various temperatures, for $H$$\parallel$$c$-axis, and for $H$$\parallel$$ab$-planes. (c) Critical current density as a function of temperature, $J_c$($T$), at self-field and at 10 kOe applied magnetic field for $H$$\parallel$$ab$-planes.}
\label{fig5}
\end{figure}

Figure 6 shows the result of the normalized pinning force density, $F_p$/$F_{pmax}$, as a function of reduced magnetic field, $h$ = $H$/$H_{irr}$, obtained for $H$$\parallel$$c$-axis and $H$$\parallel$$ab$-planes, where $F_p$ = $J_c$$\times$$H$. The  collapse of all curves in Fig. 6 evidences that one type of pinning is dominant in the sample. This allows us to fit the final collapsed curve to the well known Dew-Hughes expression \cite{dew}, $F_p$/$F_{pmax}$ = $A$$h$$^p$(1-$h$)$^q$, and extract the peak field $h_{max}$ related to the maximum pinning force predicted to occur at $h_{max}$ = $p$/($p$+$q$). The different values of $p$ and $q$ can help to determine the dominant source of pinning \cite{dew,ko,matin,shyam2015,sha}. 

\begin{figure}[htb]
\includegraphics[width=\linewidth]{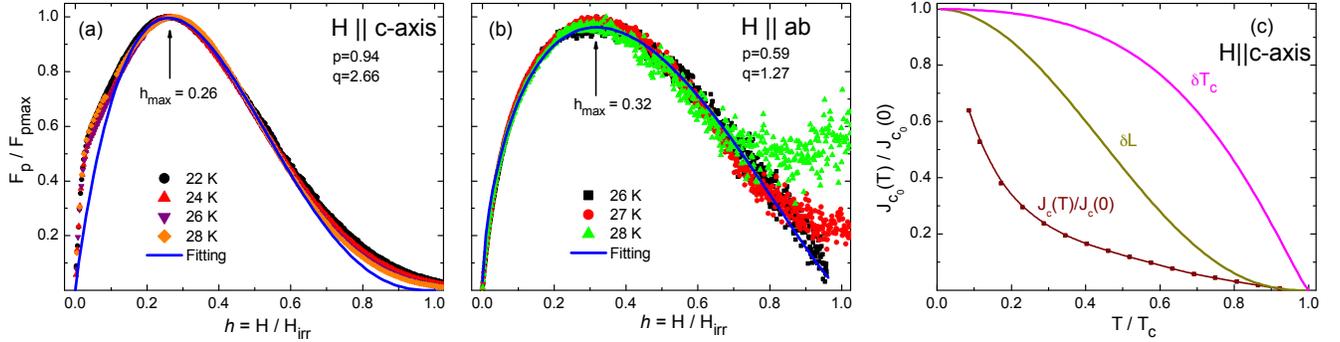}
\caption{Normalized pinning force density, $F_p$/$F_{pmax}$, as a function of reduced magnetic field, $h$ = $H$/$H_{irr}$, for (a) $H$$\parallel$$c$-axis, and for (b) $H$$\parallel$$ab$-planes, at different temperatures. Solid lines in (a,b) are fit to the data using the Dew-Hughes model (see text). Peak positions, $p$/($p+q$), obtained from fitting are consistent the $h_{max}$ realized in scaled with the $F_p$/$F_{pmax}$ curves. (c) Critical current density normalized with its zero temperature limiting value, $J_c(T)$/$J_c($T$ = 0)$, plotted as a function of reduced temperature, $T$/$T_c$, for $H$$\parallel$$c$-axis. The line joining experimental values is a guide to the eyes. Solid lines represent the temperature dependence of $J_c(T)$/$J_c($T$ = 0)$, for theoretical models that describe the pinning due to the charge carrier mean free path, $\delta$$l$, and due to the variation in superconducting transition temperature, $\delta$$T_c$.}
\label{fig6}
\end{figure}

Figure 6a shows the plot and fitting of the normalized pinning force density for $H$$\parallel$$c$-axis, which yield $p$ = 0.94 and $q$ = 2.66, where the maximum $h_{max}$ = 0.26 coincides with the expected value $h_{max}$ = $p$/($p$+$q$) = 0.26. According to the classic Dew-Hughes model, $h_{max}$ = 0.33 and 0.2 suggests the pinning due to point and surface defect respectively. Therefore, in the present case, $h_{max}$ = 0.26 is suggestive of pinning due to both point defects as well as surface defects. The observed value of $h_{max}$ as well as the values of the parameters $p$ and $q$ are quite similar to the ones obtained for BaFe$_{2-x}$Ni$_x$As$_2$ \cite{18}. From Dew Hughes work, values of $h_{max}$ smaller than 0.5 with $p$ = 1 and $q$ = 2 are due to $\delta$$l$ pinning and point pins, which appears to be the case for our sample with $H$$\parallel$$c$-axis. However, our value of $q$ is higher than 2 and values of $p$ and $q$ larger than 2 are not explained in Dew Hughes work\cite{ko}. Figure 6b shows the results and fitting of the normalized pinning force density for $H$$\parallel$$ab$-planes, with $p$ = 0.59,  $q$ = 1.27 and $h_{max}$ = 0.32 = $p$/($p$ + $q$), which suggests the dominant role of point pinning in the sample. 

Figure 6c shows the temperature dependence of the zero field critical current density, $J_c$($H$ = 0, $T$), normalized by the zero field critical current density obtained at zero temperature, for $H$$\parallel$$c$-axis. The experimental values in Fig. 6c are compared with the predicted expression for $\delta$$l$-type of pinning, $J_c(T)$/$J_c($T$ = 0)$ = (1+$t$$^2$)$^{-1/2}$(1-$t$$^2$)$^{5/2}$, and for $\delta$$T_c$-type of pinning, $J_c(T)$/$J_c($T$ = 0)$ = (1-$t$$^2$)$^{7/6}$(1+$t$$^2$)$^{5/6}$ (ref. \cite{gri}). However, the experimental data in Fig. 6c can not be fully explained using only either $\delta$$l$ or $\delta$$T_c$-type pinning. This is due to more than one type of defect sites responsible for the vortex pinning in the sample. Such behavior has already been seen in other pnictide superconductors \cite{ep5, shyam2019, Vlasenko2015}. Since $\delta$$l$-type of pinning curve is closer to the experimental values, therefore $\delta$$l$-type pinning is likely to be the dominant one in the present case.

\section*{Conclusions}

In conclusion, we observed that the anisotropic bilayer pnictide superconductor KCa$_2$Fe$_4$As$_4$F$_2$ presents a SMP on the $M(H)$ curves only for $H$$\parallel$$ab$-planes, which develops below 16 K due to the emergence of Josephson vortices. A peak observed in the temperature dependence of critical current density at the same temperature, 16 K, is interpreted due to the emergence of Josephson vortices below this temperature, which is expected for samples with large anisotropy. We also observed that the values of the first peak penetration field, $H_1$, appearing in $M(H)$ curves for $H$$\parallel$$ab$-planes at temperatures below 16 K are lower than what is expected from the behavior observed at temperatures near $T_c$. The lower values of $H_1$ occurring below 16 K evidence that the magnetic field penetrates easily when Josephson vortices form inside the sample. As the temperature drops below 24 K the value of $H_1$ in $M(H)$ curves for $H$$\parallel$$ab$-planes shows a large discontinuous increase, with the temperature behavior of these peak field, called $H_a$, smoothly joining the line formed by the temperature dependence of the SMP peak, $H_p$. A vortex dynamics study suggests that the SMP can be explained in terms of an elastic to plastic pinning crossover, while the peak $H_a$ observed between 23 K and 17 K is likely due to the disordered vortex lattice that appeared at lowered field. Interestingly, vortices in this disordered lattice show plastic pinning. Scaled pinning force curves suggest point, as well as surface defects, contribute to the pinning for $H$$\parallel$$c$-axis, whereas, for $H$$\parallel$$ab$-planes, point defects are the dominant source of pinning. The crystal shows $J_c$($H$ = 0) exceeding 10$^6$ A/cm$^2$ for temperatures below 14 K for $H$$\parallel$$c$-axis, which signals this a potential material for applications.

\section*{Methods}

The single crystal under study was grown by the self flux method \cite{crystal,77,78,79,80}. It has a mass of 0.47 mg, dimensions of 2.46$\times$1.50$\times$0.04 mm$^3$, density of 4.87 g/cm$^3$, anisotropy $\gamma$$\sim$ 15\cite{wang, pin1} near $T_c$, and a considerably sharp $T_c$ $\simeq$ 34 K as measured by zero field cooled, (ZFC), $M$($T$) with the remanent field of the magnet $H$ $\sim$ 1 Oe, applied parallel to the $c$-axis of the sample. All the magnetization measurements shown in the paper (except the data shown in fig. 1(b)) were performed using a vibrating sample magnetometer (VSM) inserted in a 9 T physical property measurements system (PPMS Quantum Design). However, the data shown in fig. 1(b) was measured using a 7 T squid-vsm magnetometer from Quantum Design. The isothermal magnetic hysteresis curves, $M(H)$, and magnetic relaxation curves, $M(t)$, were obtained for $H$ applied both parallel and perpendicular to the $ab$-plane of the sample, with the initial field increasing branch starting after the target temperature is reached in ZFC mode. 


\section*{Acknowledgements}
PVL is supported by an MSc. grant from Conselho Nacional de Desenvolvimento Científico e Tecnológico (CNPq). SS was supported by a post-doctoral fellowship from Fundação Carlos Chagas Filho de Amparo à Pesquisa do Estado do Rio de Janeiro (FAPERJ), project E-26/202.323/2021. LG was supported by FAPERJ, Projects E-26/010.001497/2019 and E-26/202.820/2018, and CNPq, project 308899/2021-0. This work is also supported by the National Key Research and Development Program of China (Grant No. 2018YFA0704200), the National Natural Science Foundation of China (Grants Nos. 11822411 and No. 11961160699), the Strategic Priority Research Program (B) of the CAS (Grants No. XDB25000000), the K. C. Wong Education Foundation (GJTD-2020-01), the Youth Innovation Promotion Association of CAS (Grant No. Y202001), the Postdoctoral Innovative Talent program(BX2021018) and the China Postdoctoral Science Foundation (2021M700250).

\section*{Author contributions statement}

HL and WH prepared the single crystal. SS and S-S conceived the experiment. PVL and LG conducted the magnetization measurements, PVL analyzed the data with inputs from SS and S-S. PVL made the figures. S-S, SS and LG contributed to the writing of the manuscript. All authors reviewed the manuscript. 

\section*{Additional information}
The datasets used and/or analyzed during the current study available from the corresponding author on reasonable request.
Correspondence and materials should be addressed to SS and S-S.

\textbf{Competing interests} 
The authors declare no competing interests. 

\end{document}